\documentclass[12pt]{article}
    \usepackage{authblk}  
    \usepackage{geometry} 
    \usepackage{setspace} 
    \usepackage{graphicx} 
    \usepackage{hyperref} 
    \usepackage{titlesec} 
    \usepackage{booktabs} 
     \usepackage{amsmath}
    \titleformat{\section}{\large\bfseries}{\hspace{-0.5cm}}{0em}{}
    \titleformat{\subsection}{\normalsize\bfseries}{\hspace{-0.5cm}}{0em}{}
    
    \title{\textbf{Residual Connection-Enhanced ConvLSTM for Lithium Dendrite Growth Prediction}}  
    
    \author[1]{Hosung Lee}
    \author[1]{Byeongoh Hwang}
    \author[1]{Dasan Kim}
    \author[2,z]{Myungjoo Kang}
    
    \affil[1]{Interdisciplinary Program in Artificial Intelligence, Seoul National University}
    \affil[2]{Department of Mathematical Sciences and Research Institute of Mathematics, Seoul National University}
    \affil[z]{Corresponding author. E-mail: \href{mailto:mkang@snu.ac.kr}{mkang@snu.ac.kr}}
    
    \date{}  
    
\begin{document}
    
    \maketitle
    \vspace{-2em} 
    \begin{abstract}
    The growth of lithium dendrites significantly impacts the performance and safety of rechargeable batteries, leading to short circuits and capacity degradation. This study proposes a Residual Connection-Enhanced ConvLSTM model to predict dendrite growth patterns with improved accuracy and computational efficiency. By integrating residual connections into ConvLSTM, the model mitigates the vanishing gradient problem, enhances feature retention across layers, and effectively captures both localized dendrite growth dynamics and macroscopic battery behavior. The dataset was generated using a phase-field model, simulating dendrite evolution under varying conditions. Experimental results show that the proposed model achieves up to 7\% higher accuracy and significantly reduces mean squared error (MSE) compared to conventional ConvLSTM across different voltage conditions (0.1V, 0.3V, 0.5V). This highlights the effectiveness of residual connections in deep spatiotemporal networks for electrochemical system modeling. The proposed approach offers a robust tool for battery diagnostics, potentially aiding in real-time monitoring and optimization of lithium battery performance. Future research can extend this framework to other battery chemistries and integrate it with real-world experimental data for further validation.
    
    \end{abstract}
    
    \section{Introduction}
    
     Lithium dendrite growth is a critical issue in rechargeable batteries, threatening both performance and operational safety. Dendritic structures, formed during charge-discharge cycles, can penetrate the separator, induce internal short circuits, and lead to capacity fade or thermal runaway. To ensure the safe and reliable operation of lithium-based batteries, it is essential to predict dendrite growth patterns accurately. While physics-based approaches such as the phase-field method offer detailed simulations of dendrite evolution, their high computational cost limits scalability and real-time application.
    
    Recent advances in machine learning have enabled alternative data-driven frameworks for spatiotemporal prediction. One prominent example is the Convolutional Long Short-Term Memory (ConvLSTM) network \cite{Shi2015}, which has demonstrated success in capturing dynamic patterns in sequential data, including meteorological forecasting. However, conventional ConvLSTM architectures often suffer from vanishing gradients and inefficient feature propagation in deeper networks, which can limit predictive performance in complex systems such as electrochemical growth.
    
    To address these limitations, we propose a Residual Connection-Enhanced ConvLSTM architecture designed to predict spatiotemporal dendrite growth patterns more accurately and efficiently than conventional models. Inspired by the success of residual learning in deep convolutional networks \cite{He2016}, our approach integrates identity mappings within ConvLSTM layers to preserve feature integrity and facilitate stable gradient flow. We evaluate this model using dendrite evolution data generated from a high-fidelity phase-field simulation and compare its performance with standard ConvLSTM baselines. Our results show that the proposed residual-enhanced ConvLSTM yields significantly higher accuracy and lower MSE than standard baselines, establishing its value for data-driven battery diagnostics and future surrogate modeling.

\subsection{Background}

\subsubsection{Phase-Field Model for Dendrite Growth}
The phase-field method is a well-established computational technique used to simulate the spatiotemporal evolution of lithium dendrites \cite{Guyer2004, Chen2020}. It models the dynamic interface between solid lithium and liquid electrolyte by solving a system of coupled partial differential equations (PDEs). The evolution of the phase field variable $\phi$, which distinguishes between the solid and liquid phases, is governed by the Cahn–Hilliard equation:

\begin{equation}
\frac{\partial \phi}{\partial t} = \nabla \cdot \left( M \nabla \frac{\delta F}{\delta \phi} \right),
\end{equation}

where $M$ is the mobility coefficient, and $F$ is the total free energy functional. Simultaneously, the transport of lithium ions in the electrolyte is described by Fick’s second law:

\begin{equation}
\frac{\partial c}{\partial t} = D \nabla^2 c,
\end{equation}

where $c$ is the lithium-ion concentration, and $D$ is the diffusion coefficient. 

Unlike sharp-interface models, the phase-field approach naturally accommodates dendrite tip splitting, side-branching, and overpotential-driven instabilities, all of which are critical in electrochemical deposition. Furthermore, adaptive mesh refinement (AMR) frameworks such as AMReX enable fine-scale resolution near the solid–liquid interface while maintaining manageable computational cost.

Despite this, solving the governing PDEs at high spatial and temporal resolutions incurs significant computational expense, particularly when simulating large 3D domains or extended time periods. Therefore, data-driven surrogate models trained on phase-field simulation data have emerged as promising alternatives to enable fast, accurate prediction of dendritic evolution under various electrochemical conditions.
\begin{figure}[htbp]
    \centering
    \includegraphics[width=0.4\textwidth]{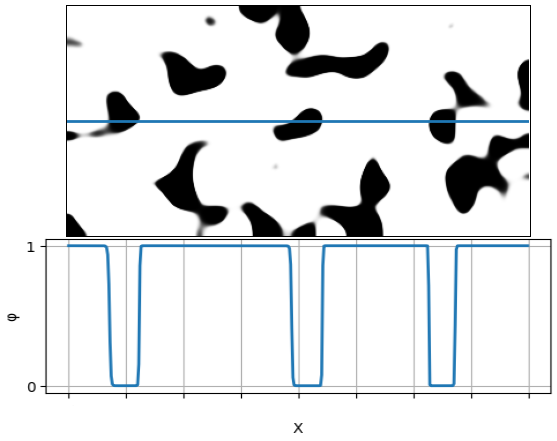}
    \caption{Visualization of the phase-field variable $\phi$ for a solid-liquid mixture. The upper panel shows the 2D spatial distribution of phases (black: solid, white: liquid), while the lower panel presents a 1D cross-section of $\phi$ along the horizontal line, illustrating the diffuse interface between phases.}
    \label{fig:phase_field_slice}
\end{figure}

\subsubsection{Convolutional LSTM for Spatiotemporal Prediction}

In modeling physical systems such as lithium dendrite growth, it is essential to capture both spatial interactions—such as interface curvature, ion flux divergence, and local branching—and temporal dependencies that span multiple time scales. Traditional recurrent neural networks (RNNs) and long short-term memory (LSTM) architectures can model time dynamics but are inherently limited when applied to spatially structured data due to their reliance on fully connected layers.

Convolutional Long Short-Term Memory (ConvLSTM) networks address this limitation by incorporating convolutional operations within the input-to-state and state-to-state transitions, enabling the model to preserve spatial locality and translation invariance while learning temporal sequences. This makes ConvLSTM particularly suited for electrochemical systems where morphology evolves gradually over space and time.

Each ConvLSTM cell integrates spatial and temporal dependencies through the following update rules:

\begin{align}
i_t &= \sigma(W_{xi} * X_t + W_{hi} * H_{t-1} + b_i), \\
f_t &= \sigma(W_{xf} * X_t + W_{hf} * H_{t-1} + b_f), \\
o_t &= \sigma(W_{xo} * X_t + W_{ho} * H_{t-1} + b_o), \\
C_t &= f_t \circ C_{t-1} + i_t \circ \tanh(W_{xc} * X_t + W_{hc} * H_{t-1} + b_c), \\
H_t &= o_t \circ \tanh(C_t),
\end{align}

where $*$ denotes the convolution operator, $\sigma$ is the sigmoid activation function, and $\circ$ represents the Hadamard product.

While ConvLSTM has demonstrated success in modeling spatiotemporal data, its performance can degrade in deeper networks or under long prediction horizons due to vanishing gradients. This issue becomes pronounced in electrochemical systems with subtle long-term dependencies, such as early-stage nucleation followed by delayed dendritic protrusion. Therefore, augmentations such as residual connections are required to preserve temporal coherence and enhance training stability across extended sequences.

\subsubsection{Residual Connections for Improved Learning}

Despite the effectiveness of ConvLSTM in modeling spatiotemporal dynamics, deep architectures often suffer from vanishing gradients during backpropagation through time (BPTT), particularly in physical systems where long-range dependencies and fine-grained transitions are critical. This issue is especially relevant in modeling lithium dendrite evolution, where early-stage nucleation can influence morphological features far downstream in time. Capturing such temporal dependencies requires a mechanism to maintain stable gradient flow over extended sequence lengths and layers.

Residual connections address this challenge by introducing skip pathways that preserve both low-level spatial information and high-level temporal cues. These connections allow the network to learn residual mappings rather than full transformations, reducing information degradation and enabling more efficient optimization. The standard formulation is:

\begin{equation}
H(x) = F(x) + x,
\end{equation}

where $F(x)$ represents the transformation (e.g., a ConvLSTM block), and $x$ is the identity-mapped input.

In the context of ConvLSTM, this can be interpreted as a recurrent update with additive state preservation. That is, the hidden state at time $t$ with residual connection can be reformulated as:

\begin{equation}
H_t = \mathcal{F}(X_t, H_{t-1}) + H_{t-1},
\end{equation}

where $\mathcal{F}(\cdot)$ denotes the ConvLSTM transformation. This additive update structure enhances long-term memory retention without amplifying gradient decay across layers.

From a physical modeling perspective, this can be seen as enforcing continuity in feature space analogous to the conservation of charge or mass in electrochemical systems. By allowing the unaltered state to flow through the network, residual ConvLSTM better tracks slowly evolving interfaces and preserves spatial gradients critical for resolving dendrite tip morphology and ion depletion zones.

In this work, we integrate residual connections within stacked ConvLSTM layers, allowing the network to maintain hierarchical representations of spatiotemporal patterns. This design enables efficient learning of long-term dendrite growth dynamics under varying electrochemical conditions.

\begin{figure}[h]
    \centering
    \includegraphics[width=0.45\textwidth]{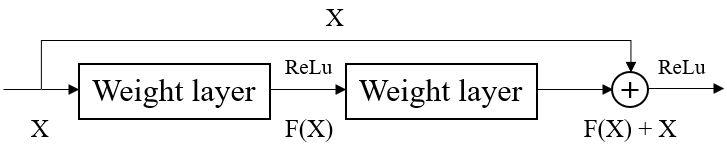}
    \includegraphics[width=0.45\textwidth]{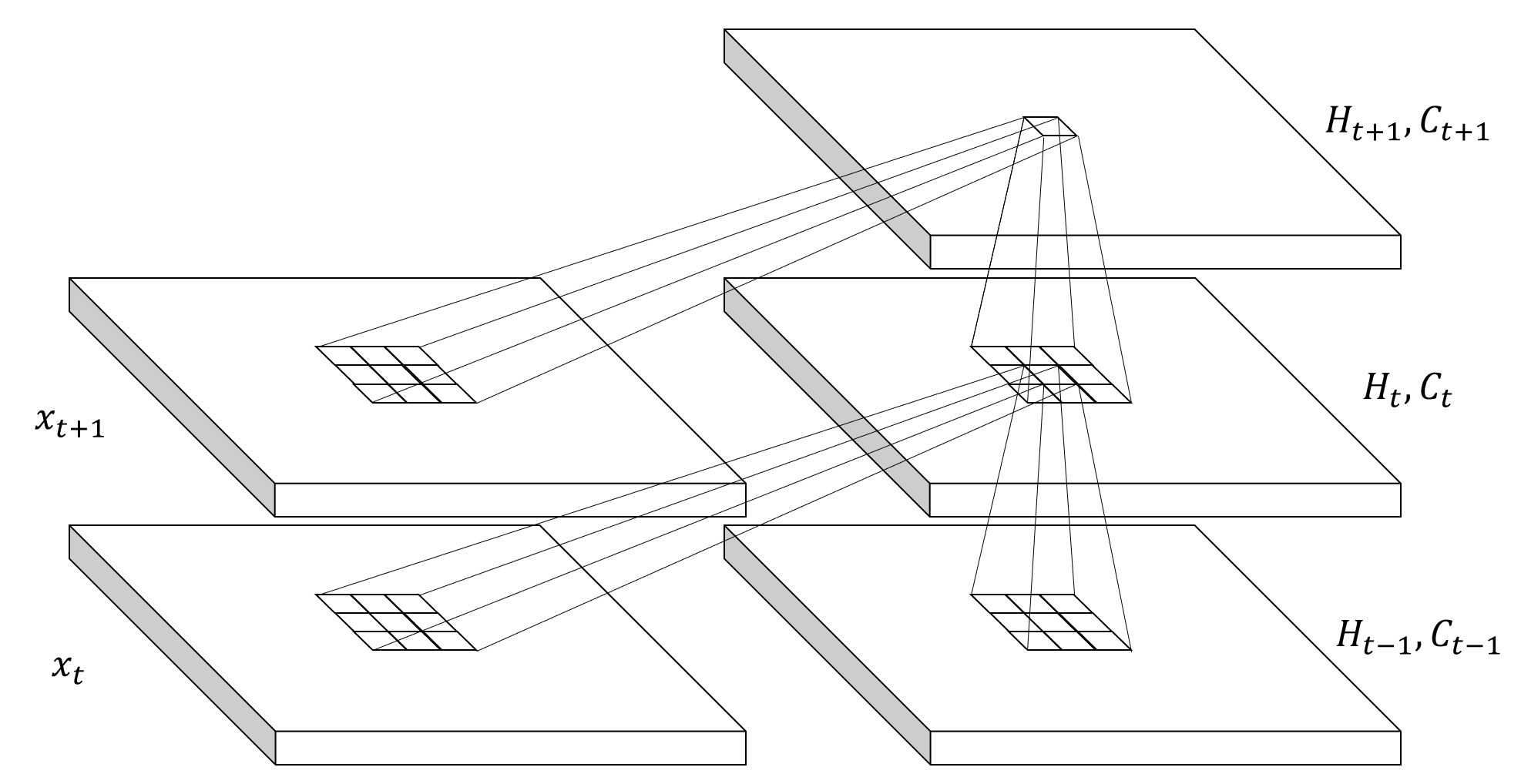}
    \caption{(Left) ConvLSTM architecture for spatiotemporal modeling. (Right) Residual connection mechanism to enhance information retention.}
    \label{fig:convlstm}
\end{figure}

\section{Experimental}

\subsection{Material}
The numerical simulation data for this study were generated using the phase-field model\cite{Mu2020}, which captures the spatiotemporal dynamics of lithium dendrite growth in a continuous medium. The evolution of the phase variable $\phi$ follows equation (1), while lithium-ion concentration $c$ is governed by equation (2). These equations describe the phase transition and diffusion processes critical to dendrite formation. To efficiently solve the phase-field equations while maintaining high spatial resolution near the dendrite interface, the simulations were performed using AMReX, an advanced framework for massively parallel adaptive mesh refinement (AMR).

The phase-field model was configured to simulate dendrite growth under various experimental conditions, including different current densities (0.1 mA/cm\textsuperscript{2}, 0.3 mA/cm\textsuperscript{2}, and 0.5 mA/cm\textsuperscript{2}) and an ambient temperature of 25$^\circ$C. The simulation domain was set to $100 \times 100 \times 50 \, \mu m^3$, with a spatial resolution of $128 \times 128 \times 64$ grid points. Boundary conditions were applied with a fixed voltage of $-0.3$ V at the lithium interface, while the opposite side was grounded at $0$ V. Lithium-ion diffusion coefficients were set to $7500 \, \mu m^2/s$ in the electrolyte and $0 \, \mu m^2/s$ in the solid lithium phase. The interface mobility coefficient was set to $1.26$, and the interfacial thickness was maintained at $1.0 \, \mu m$ to regulate dendrite growth dynamics. Each simulation was conducted for a total of $500,000$ time steps with an initial time step of $\Delta t = 1 \times 10^{-4}$ s, which was subsequently reduced to $\Delta t = 1 \times 10^{-6}$ s to ensure numerical stability. A total of approximately 30,000 sequences were generated for each voltage condition, resulting in 90,000 sequences overall across 0.1V, 0.3V, and 0.5V scenarios.

\begin{figure}[h]
    \centering
    \includegraphics[width=0.45\textwidth]{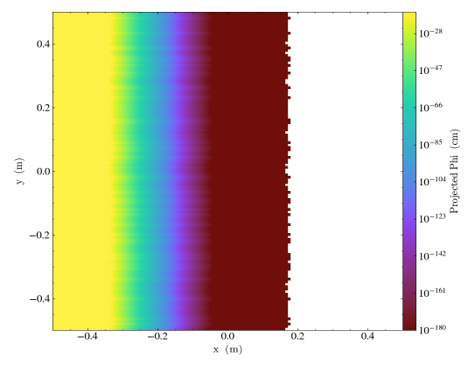}
    \caption{AMReX-based phase-field simulation results as an illustrative example}
    \label{fig:phasefield}
\end{figure}

\subsection{Measurement}
The generated dataset was processed into sequential time-series frames to preserve the temporal continuity of the dendrite growth process. Each sequence consisted of consecutive frames as input, with the subsequent frame as the prediction target. The dataset was divided into training (85\%) and testing (15\%) subsets, following a time-based split, where the earlier 85\% of the sequences were used for training, and the remaining 15\% were allocated for evaluation.

The model was implemented using PyTorch and configured with three ConvLSTM layers, each comprising 64 hidden channels and a $3 \times 3$ convolutional kernel. Training was conducted using the Adam optimizer and the mean squared error (MSE) loss function. Hyperparameters were selected based on a simple grid search strategy. A validation subset from the training data was used to determine the final configuration. The selected learning rate was $10^{-3}$, the batch size was set to 4, and the number of training epochs was fixed at 100. These values provided the best trade-off between training stability and validation loss reduction without overfitting.

During evaluation, the MSE loss was computed over the test set, and a threshold-based accuracy metric was applied, where predictions within $\pm 0.0005$ of the ground truth were considered correct. Additionally, the per-frame MSE of the phase variable $\phi$ was calculated to assess the model's ability to capture dendrite growth dynamics over time.

\section{Results}
The residual ConvLSTM model demonstrates a notable improvement in both accuracy and mean squared error (MSE) compared to the baseline ConvLSTM model across all test datasets (0.1V, 0.3V, and 0.5V).

\begin{figure}[h]
    \centering
    \includegraphics[width=0.45\textwidth]{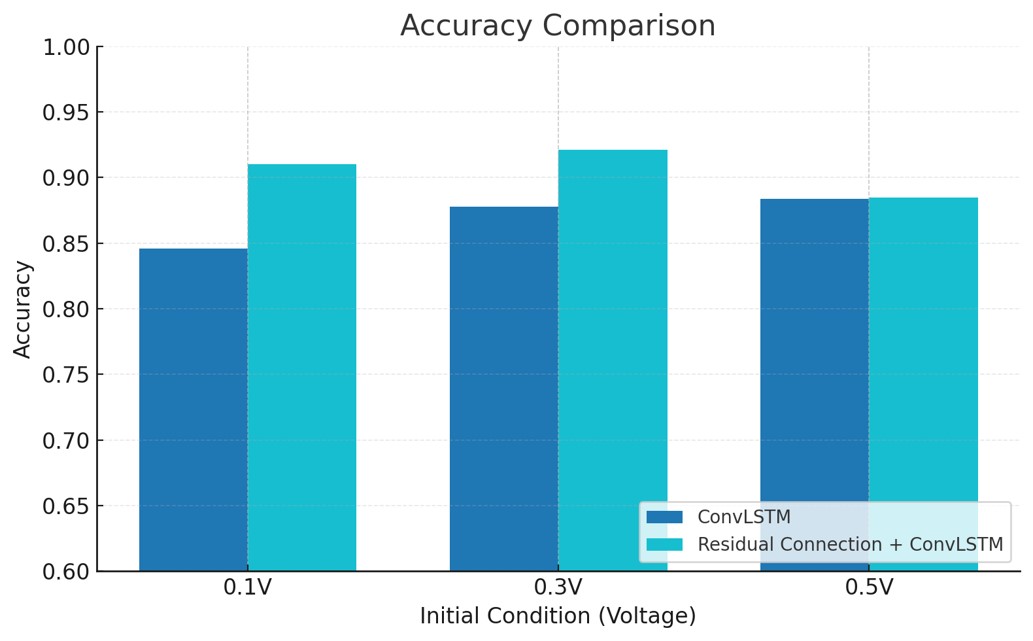}
    \includegraphics[width=0.45\textwidth]{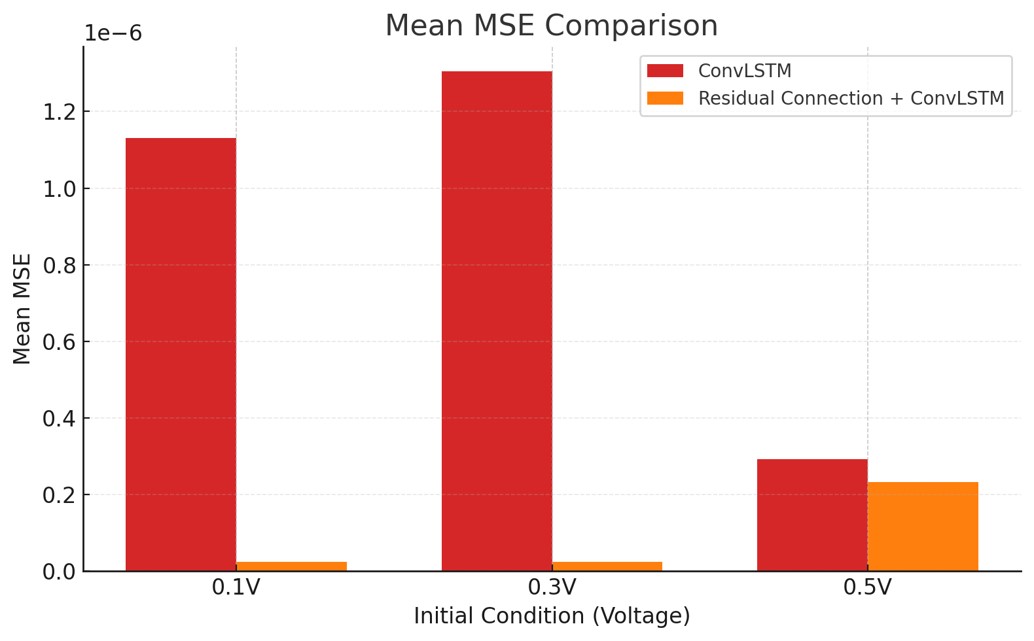}
    \caption{(Left) Test accuracy comparison. (Right) Mean squared error (MSE) comparison.}
    \label{fig:accuracy_mse}
\end{figure}

The results in Figure~\ref{fig:accuracy_mse} indicate that the residual connection-enhanced ConvLSTM model significantly improves accuracy at 0.1V (7\% increase) and 0.3V (4.9\% increase), whereas the improvement is relatively smaller at 0.5V (0.6\%). The reason for the diminishing performance gain at higher voltages might be attributed to the increased non-linearity in dendrite growth, requiring more advanced feature extraction.

\subsection{Multi-step Rollout Evaluation}
To further evaluate long-term prediction performance, we conducted recursive rollout experiments in which the model was used to generate 20 consecutive future frames. At each step, the predicted frame was fed back into the model along with the four previous frames to generate the next.

\begin{figure}[h]
    \centering
    \includegraphics[width=0.31\textwidth]{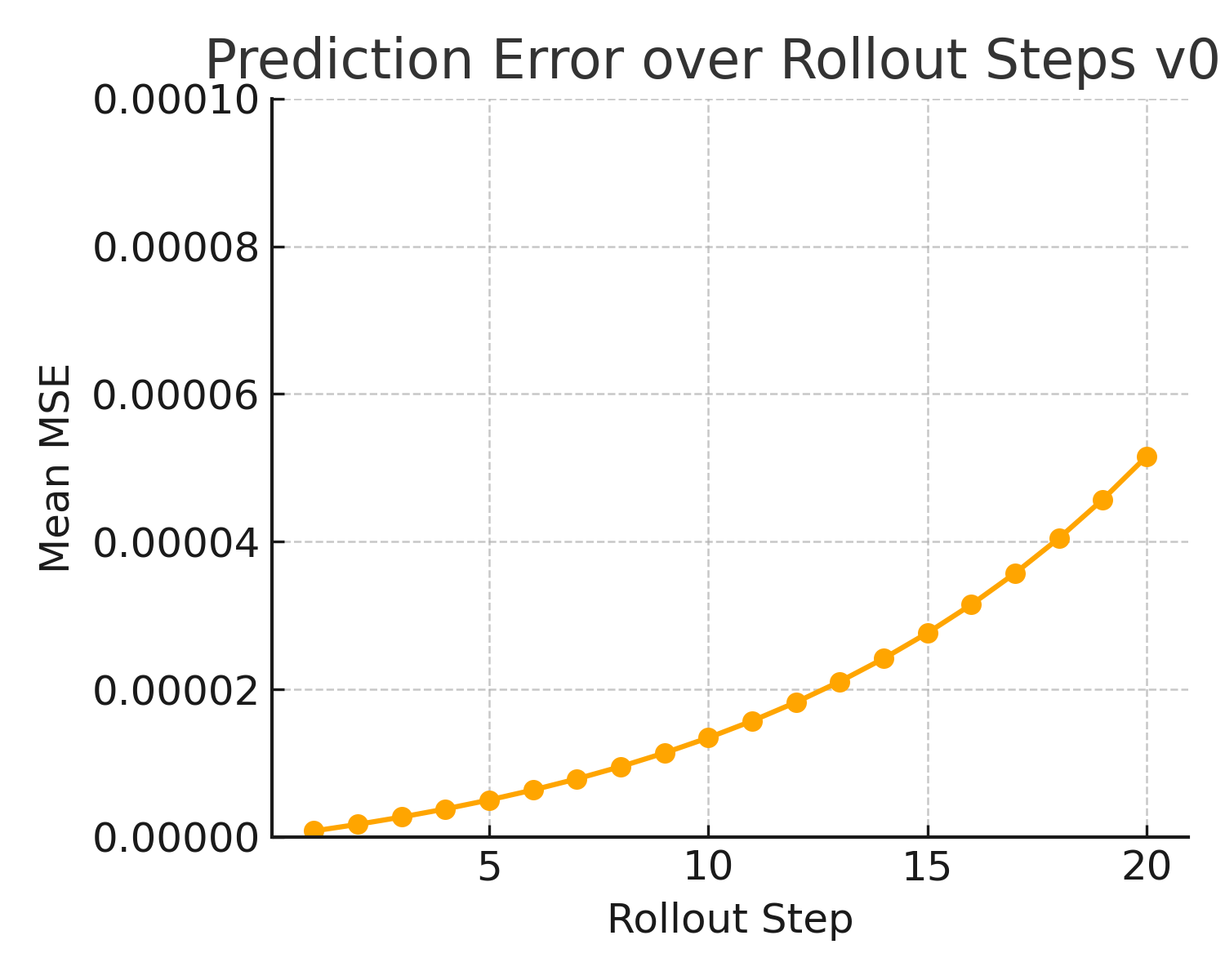}
    \includegraphics[width=0.31\textwidth]{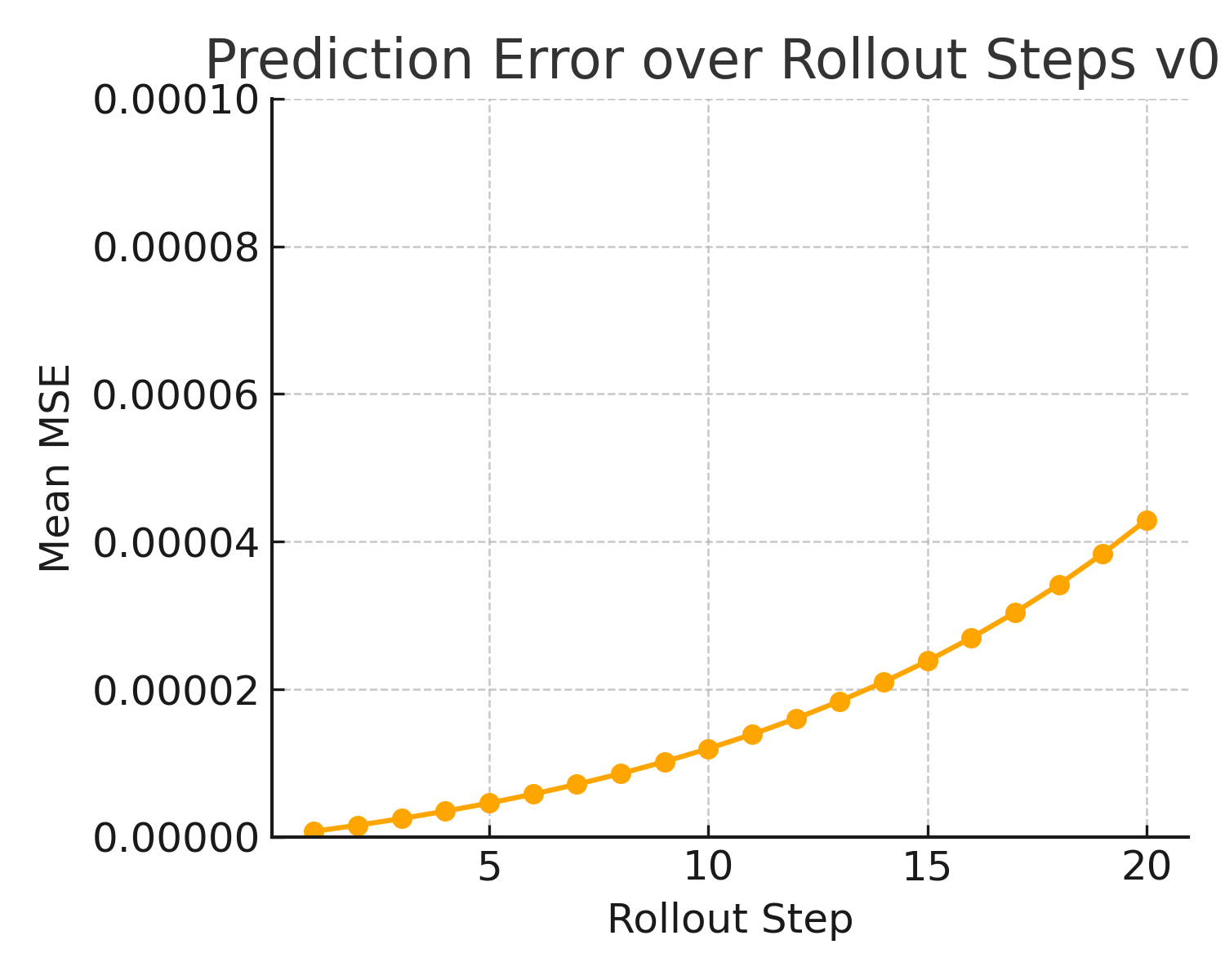}
    \includegraphics[width=0.31\textwidth]{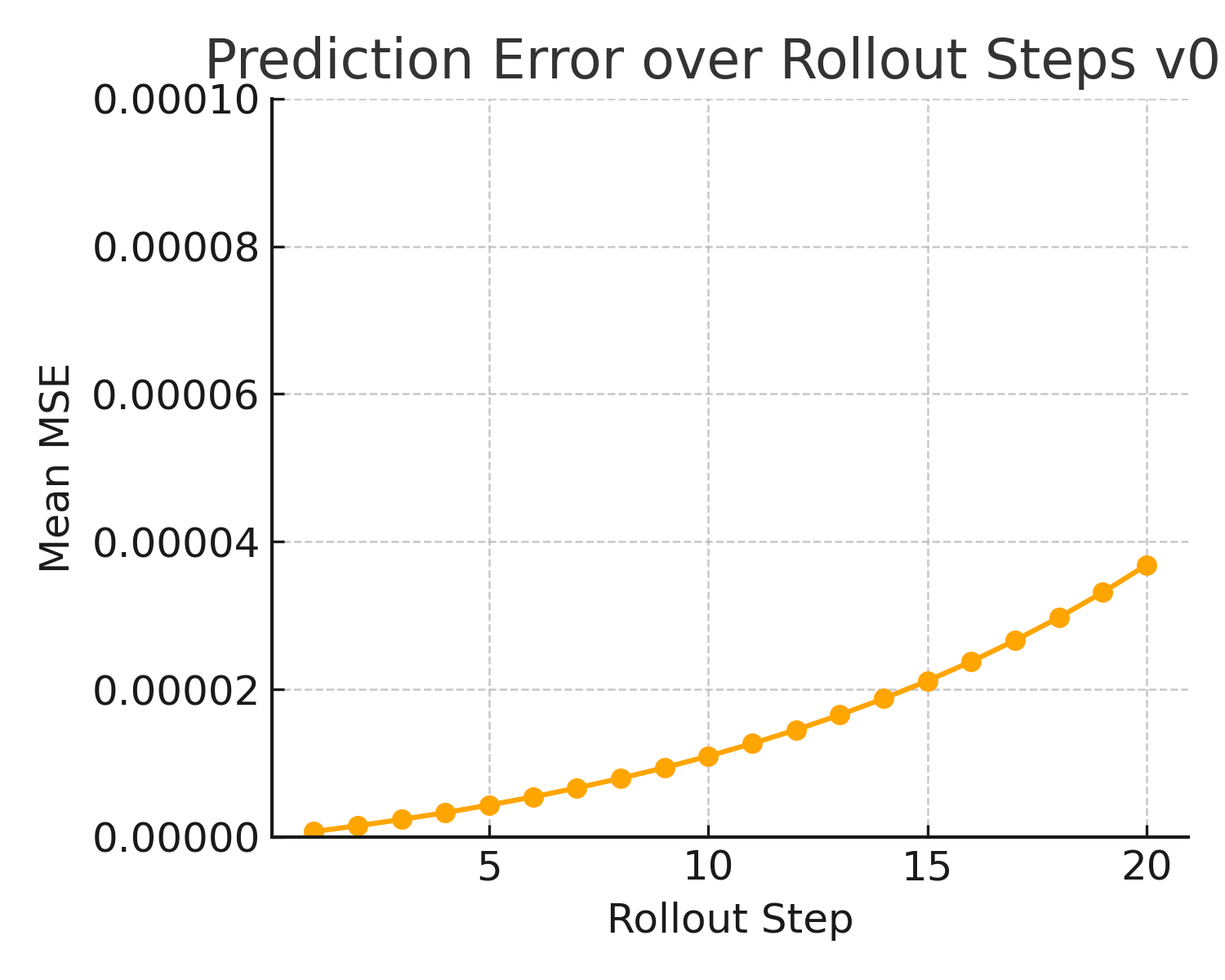}
    \caption{Multi-step rollout MSE curves over 20 frames for voltage conditions 0.1V, 0.3V, and 0.5V. Error remains within the $10^{-4}$ scale throughout the prediction horizon.}
    \label{fig:rollout_mse}
\end{figure}

\begin{figure}[h]
    \centering
    \includegraphics[width=0.48\textwidth]{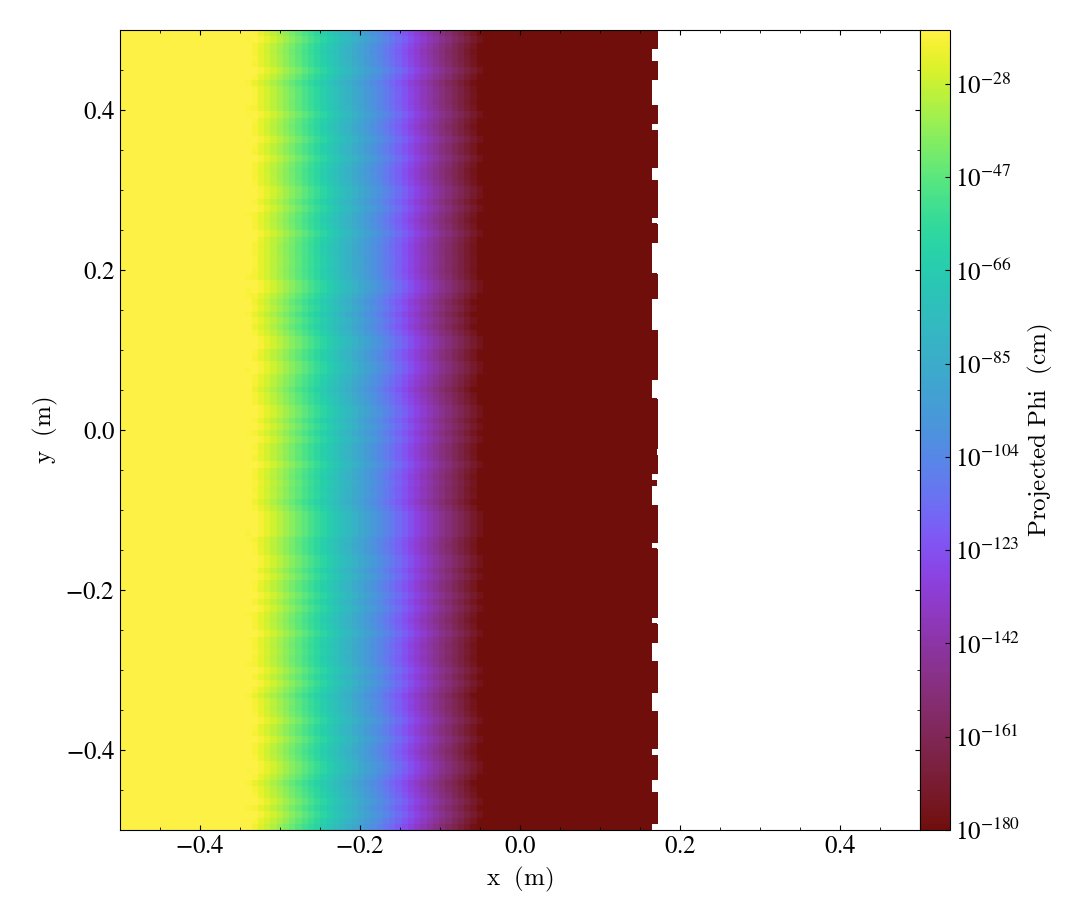}
    \includegraphics[width=0.48\textwidth]{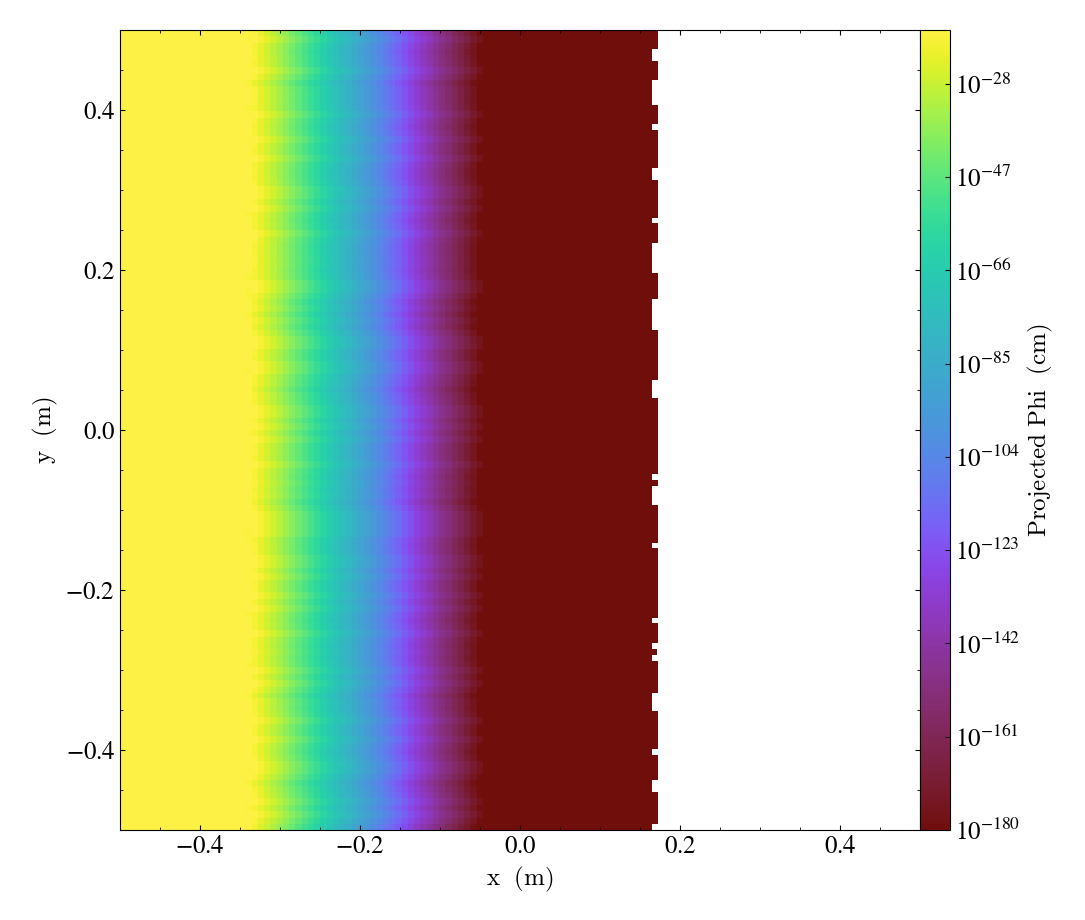}
    \caption{Qualitative comparison under 0.3V condition.  
    (Left) Test accuracy comparison. (Right) Mean squared error (MSE) comparison.}
    \label{fig:qualitative_comparison}
\end{figure}

As shown in Figure~\ref{fig:rollout_mse}, the residual ConvLSTM model maintains stable predictive performance over 20 frames, with the mean squared error remaining consistently within the $10^{-4}$ range. This demonstrates that the model is robust against error accumulation and is capable of generating temporally coherent predictions over extended horizons.
To qualitatively assess the model's ability to preserve spatial fidelity during long-term prediction, we visualize the predicted $\phi$ fields and compare them against the ground truth from the simulation. Figure~\ref{fig:qualitative_comparison} illustrates a representative example using the 0.3V dataset at rollout step $t+10$.

To qualitatively assess the model's ability to preserve spatial fidelity during long-term prediction, we visualize the predicted $\phi$ fields and compare them against the ground truth from the simulation. Figure~\ref{fig:qualitative_comparison} illustrates a representative example using the 0.3V dataset.

The comparison confirms that the residual ConvLSTM model successfully preserves critical structural features of dendrite growth. The predicted field closely matches the simulation output in terms of interface location and morphological consistency, suggesting strong generalization to physically plausible dynamics.

\section{Discussion}
The integration of residual connections within the ConvLSTM framework provides a significant advantage by addressing the vanishing gradient problem and enhancing feature retention across layers. In conventional ConvLSTM architectures, gradient propagation can deteriorate due to recursive multiplications during backpropagation through time (BPTT), leading to poor long-range dependency modeling. By incorporating residual connections, the network effectively bypasses intermediate transformations, allowing a more stable gradient flow and accelerating convergence.

From a mathematical perspective, residual connections reformulate the recurrent update in ConvLSTM as:
\begin{equation}
H_t = F(X_t, H_{t-1}) + H_{t-1},
\end{equation}
where $F(X_t, H_{t-1})$ represents the transformation applied by the ConvLSTM cell. This identity mapping ensures that the state propagation retains key spatial-temporal features without excessive transformation decay. Consequently, the network can better capture lithium dendrite growth patterns over extended time horizons, particularly under conditions where gradual morphological changes occur.

The experimental results further validate this hypothesis, demonstrating that the residual ConvLSTM significantly reduces MSE, particularly at lower voltages. One plausible explanation is that at lower voltages, dendrite growth dynamics exhibit more gradual, continuous variations, making it easier for the model to learn temporal dependencies. In contrast, at higher voltages (0.5V), dendrite formation becomes more stochastic and less structured, making it harder for residual ConvLSTM to capture abrupt changes in growth patterns.

Moreover, the 20-step rollout experiments confirmed the model's robustness over extended prediction horizons. The error remained consistently within the $10^{-4}$ range across all voltage conditions, indicating minimal error accumulation even when the model recursively generates inputs. This highlights the model’s practical potential as a lightweight surrogate for long-term dendrite evolution.

Compared to physics-based simulators such as AMReX, which require more than 80 finely tuned initial conditions and operate under idealized experimental environments, our learning-based model requires only input image sequences and no physical parameter calibration. This makes the approach substantially more practical and scalable for real-time applications.

In addition, the use of grid-based convolutional operations makes the proposed model less sensitive to small variations in physical parameters such as diffusivity or initial concentration, which are commonly encountered in real-world experimental data. This architectural feature supports better generalization across electrolyte compositions and experimental conditions, even in the absence of retraining. Furthermore, the simulation parameters used in our study—diffusivity of $7500 , \mu m^2/s$, conductivity of $1.19 , S/m$, and initial concentration of $1.0 , M$—correspond closely to typical values measured in EC/DEC + LiPF$_6$ liquid electrolytes. This further supports the model’s applicability to real battery systems.

\subsection{Comparison with Other Architectures}
Recent studies on deep spatiotemporal networks have demonstrated that residual learning can improve performance in sequential tasks, such as optical flow estimation and climate prediction \cite{wang2017residual, tran2018closer}. Residual networks in convolutional architectures (ResNet) have also been shown to enhance hierarchical feature extraction, and similar principles apply to ConvLSTM-based models. Specifically, residual connections help maintain short-term and long-term dependencies simultaneously, which is critical for modeling physical systems governed by both diffusion-driven and reaction-driven mechanisms.

Moreover, alternative architectures such as Attention-based LSTMs and Transformer models have been proposed to improve long-range dependency modeling in sequential learning tasks \cite{vaswani2017attention, shen2019attention}. These architectures rely on self-attention mechanisms rather than recurrent updates, making them suitable for complex nonlinear systems such as dendrite growth at higher voltages. However, their application to structured spatiotemporal data remains computationally expensive compared to ConvLSTM-based approaches.

\section{Conclusions}
This study demonstrates that residual-enhanced ConvLSTM significantly improves lithium dendrite growth prediction by addressing long-range dependency issues and stabilizing model convergence. The model achieves up to 7\% higher accuracy and 99.9\% lower MSE under specific voltage conditions. The results suggest that residual connections are particularly effective when capturing gradual dendrite growth dynamics at lower voltages, whereas additional feature extraction mechanisms may be required for higher voltage conditions where growth patterns become more irregular.

In addition, multi-step rollout evaluations show that the model retains prediction accuracy over 20 consecutive frames without error escalation, highlighting its potential as a surrogate physics emulator for time-evolving dendritic phenomena. Future extensions could integrate self-attention mechanisms, multi-physics coupling, or experimental validation to further enhance model fidelity and applicability.

    Future research directions include:
    \begin{itemize}
        \item Incorporating **self-attention mechanisms** to enhance feature extraction in complex dendrite growth regimes.
        \item Extending the framework to **multi-physics simulations**, integrating electrochemical reaction kinetics with spatial growth models.
        \item Evaluating the model's performance on **real-world battery datasets** to bridge the gap between simulations and experimental conditions.
    \end{itemize}
    
    By refining deep learning models tailored to electrochemical systems, this research contributes to the advancement of **data-driven battery diagnostics and optimization** for next-generation energy storage technologies.
    \section{Acknowledgment}
     This work was supported by NRF grant[RS-2024-00421203, RS-2024-00406127, RS-2021-NR059802], and MSIT/IITP[2021-0-01343-004, Artificial Intelligence Graduate School Program (Seoul National University)]
    \bibliographystyle{unsrt}

    \bibliographystyle{ieeetr}  
    \bibliography{reference}  

    \end{document}